\documentclass{aa}
\usepackage{graphics}
\begin{document}

   \thesaurus{03 
              (02.19.2;  
               08.19.5;  
               12.04.3)} 

\title{ A Search for Candidate Light Echoes: Photometry of
Supernova Environments
\thanks{Tables 2a and 2b are also available in
electronic form and figures 1 through 36 are only available in
electronic form at the CDS via anonymous ftp to cdsarc.u-strasbg.fr
(130.79.128.5) or via http://cdsweb.u-strasbg.fr/Abstract.html.}}

\author{Francesca R. Boffi, William B. Sparks,
F. Duccio Macchetto\thanks{Affiliated to the Astrophysics Division of the Space
Science Department of ESA}}

\offprints{F.R. Boffi}

\institute{Space Telescope Science Institute\\
3700 San Martin Drive, \\
Baltimore, MD, 21218, USA\\
email: boffi@stsci.edu, sparks@stsci.edu, macchetto@stsci.edu}

\date{Received April 19, 1999; Accepted June 7, 1999}

\titlerunning{Supernova Light Echoes}
\authorrunning{Boffi et al.}
\maketitle

\begin{abstract}

Supernova (SN) light echoes could be a powerful tool for determining
distances to galaxies geometrically, Sparks 1994. In this paper we
present CCD photometry of the environments of 64 historical supernovae,
the first results of a program designed to search for light echoes from
these SNe.

We commonly find patches of optical emission at, or close to, the sites
of the supernovae. The color distribution of these patches is broad,
and generally consistent with stellar population colors, possibly with
some reddening. However there are in addition patches with both
unusually red and unusually blue colors.  We expect light echoes to be
blue, and while none of the objects are quite as blue in V-R as the
known light echo of SN1991T, there are features that are unusually blue
and we identify these as candidate light echoes for follow-on observations.

\keywords{Scattering -  supernovae: general -  distance scale}


\end{abstract}

\section{Introduction}

It is of paramount importance to determine distances to external galaxies: 
such knowledge impacts both stellar and extragalactic
astrophysics as well as cosmology and the structure of the Universe
including the derivation of the Hubble constant (${\rm H_0}$).
Unfortunately the standard ``lumi\-no\-sity-calibrated'' methods of distance
determination have not yet allowed an unambiguous determination of the
Hubble constant (see the Proceedings of ``The
Extragalactic Distance Scale'' STScI May 1996 Symposium for an
overview), and all luminosity based indicators are extinction and
metallicity dependent in some way.
Thus (new) physical and geometric methods of distance determination
are highly desirable and should be pursued wherever possible.

Sparks (1994) proposed that distances to external galaxies (well in excess
of the distance to the Virgo cluster) could be determined by means of
supernova (SN) light echoes. The method is appealing as it
is purely geometrical, does not need any secondary
distance indicators or calibration and might be used to relate to luminosity
distance indicators when applied to galaxies
hosting Cepheids and/or Type Ia supernovae (Sparks 1994, 1996).

The technique requires high resolution imaging polarization observations of
SN light echoes.
A light echo is expected to be produced by
scattering of the SN light by dust in the interstellar medium,
and would be visible to the observer at some time after
the supernova explosion (due to light travel time effects).
The evolution of such a feature is well understood
(e.g. Chevalier 1986) and mathematically straightforward.
To summarize, the light echo is a paraboloid at the focus of
which the historical supernova lies. The observer looks down the axis,
``into'' the paraboloid. While the {\it intensity} distribution of the echo
is likely to be complex, depending primarily on dust density, by contrast,
the {\it polarization} distribution should be simple. Angular distance
from the supernova is related to the scattering angle, and polarization
depends simply on scattering angle, maximizing at $90^{\circ}$ scattering.
Hence at the intersection between the plane of the sky containing the
supernova and the parabola, light is scattered at an
angle of $90$ degrees and forms a ring in an image of the degree of
polarization (this light having the maximum degree of polarization).
The linear diameter of the ring
is $\rm 2ct$, where $\rm c$ is the speed of light and
$\rm t$ is the time since the explosion, therefore,
if ${\rm {\phi}}$, the angular diameter of the ring is measured,
the distance ${\rm D}$ is derived geometrically by ${\rm D}={\rm 2ct}/{\rm
{\phi}}$.


Light echoes have been observed around SN 1987A (Crotts 1988, Sparks et al.
1989) and SN 1991T (Schmidt et al. 1994).
From the ground we would not expect to resolve
light echoes around known historical supernovae in most cases.
Yet from the ground {\it candidate} light echoes
can be sought and identified by using imaging and, when appropriate,
polarization criteria:
a good light echo candidate shows optical emission at/near the
site of the historical supernova, is blue in color,
because of both the intrinsic ``blueness'' of the SN and the additional
blueing introduced by the scattering process, and is polarized.
The discovery of light echo candidates and subsequent confirmation either
through ground--based spectroscopic
observations, or direct  space imaging polarimetry observations is an essential
step in the process of enabling general geometric galaxy distance
determinations by this technique. Here we describe a program to search
for candidate echoes via CCD imaging. Sparks et al. (1999) show how
space imaging polarimetry can be used in the well known case of the
SN 1991T light echo. The region of maximum linear polarization
emission cannot be resolved yet due to the small time
since the explosion and the distance to the galaxy. However,
we give a distance estimate to SN 1991T (an upper limit of 15 Mpc
is found) via simple modeling of the scattering process.

Although the main objective of such a search program is to find light echoes
and use them to determine distances, the photometric and
spectroscopic data collected, together with the polarimetry, should also
provide important information on the supernova environments in the
host galaxy.
More generally, what we learn about environments of different SN Types
in turn tells us something about the SN progenitor and the stellar
populations of galaxies.
In particular we think of SNe of Type Ia
whose star progenitor systems have not unambiguously been identified yet.
A careful investigation of both circumstellar and
interstellar environments of these SNe may provide
clues on the nature of their progenitors and in some instances
allow to test some of the proposed pre-SN scenarios (Boffi \& Branch 1995,
Branch et al. 1995).
The identification of Type Ia SN star system progenitors
is important (among other reasons) because the nature of the progenitors
is connected to the use of these SNe as distance indicators
and to derive cosmological parameters (Perl\-mut\-ter et al. 1998,
Riess et al. 1998).
In fact to determine the
mass and energy density of the Universe from SNe of Type Ia in the
Hubble diagram, the evolution with cosmic epoch of both the Type Ia SN rate
and their luminosity function are relevant; these functions
depend on the nature of the progenitor systems.

In Section 2 the sample of galaxies is presented and we explain
the selection criteria used. The observations are described and
the analysis process is presented in Section 3.
Our results are summarized in Section 4 where we discuss
(a) the results from our search for candidate light echoes;
(b) some by-products of the present investigation in terms of
SN environments and of SN star progenitors. Conclusions and plans
for future work are discussed in the last Section.

\section{Observations} 

\subsection{Sample Selection}

Firstly all galaxies have been selected to be
parent galaxies of historical supernovae. Many
host multiple SN events. Second, echoes, if present, would be resolved
with high resolution imaging using the Hubble Space Telescope instrumentation,
i.e. their angular size would be $\ge {0.1}$ arcsec on the long distance
scale (i.e., for a low value of the Hubble's constant).
This resulted in a total sample of
(North \& South) 172 galaxies, 211 supernovae. Due to time, visibility and
weather constraints, of these we observed 38 galaxies, hosting 64 supernovae.
Furthermore, for some, their distance
is already well determined by means of normal Type Ia supernovae and Cepheid
variable stars (e.g. NGC 5253, IC 4182), 
thus they are are already tied to traditional methods
of distance determinations and can be used for comparisons
between different methods. Priority was given to such galaxies
and to galaxies hosting multiple events. In other regards, our resulting
sample is unbiassed and representative of the larger sample.

\subsection{Observing runs}

The data presented in this paper were obtained in three observing runs:
one at the 1m Jacobus Kapteyn Telescope (JKT) in La Palma,
Canary Islands, in October 1994 and two with EFOSC, at the 3.6 m Cassegrain
telescope at the European Southern Observatory, La Silla, Chile,
respectively in March and May 1995 (see Tables
1a and 1b). In all runs B, V and R images were taken for each target.
Seventeen galaxies were observed with the
1m JKT and a further 26 with the 3.6m telescope of ESO.
All galaxies are listed, in order of increasing Right Ascension, in Tables
1a and 1b, respectively for the JKT and the ESO runs.
Both tables are structured as follows: the galaxy name (NGC,
UGC, IC, M) is given in the first column, Right Ascension and
declination follow, then the morphological Type of the galaxy is
listed in column 4. The supernova(e) is (are) given in the fifth column,
followed by their offset with respect to the nucleus of the galaxy
(unless otherwise noted);
finally host galaxy recession velocity, date of observation,
exposure times in B, V and R, respectively, are given (from column 8 through
11).
In Table 1b a letter ``p'' next to the exposure time indicates that
polarization measurements were also taken. The polarization results
will be presented elsewhere.
The B and V bands are in the Johnson system, the R in the Cousins.


For the JKT observations, only imaging observations were performed.
A TeK4 CCD detector was used.
This is a high--resolution chip
with $\rm {1024} \times {1024}$ pixels corresponding to a
$\rm {0.33}~ arcsec ~pixel^{-1}$ image scale.
The CCD was used with a gain corresponding to $\rm {0.78}$ electrons per
CCD analog--to--digital unit (adu), and the readout noise estimated from the
variance of intensity values in the overscan region was measured to be
$\rm \sim {4.7}$ electrons. Seeing was measured to be
$\sim 1.''5$ FWHM typically. Bias (zero--exposures) and flat field
frames were taken at the beginning
and end of each observing night.

For the ESO observations, both imaging and imaging
polarization observations were obtained.
We mounted four optical--quality polarizing filters
utilizing HN38 Polaroid and anti-reflection coated $\rm Mg{F_2}$ substrates
at position angles $\rm 0^\circ$ $\rm {45}^\circ$ $\rm {90}^\circ$
$\rm {135}^\circ$ in the first filter wheel to be encountered by the light beam
after passing through the aperture plate.
The ESO CCD No. 26 was used as the detector.
This is a high resolution TeK detector with $\rm {512} \times {512}$ 
pixels corresponding to a $\rm {0.61} ~arcsec ~pixel^{-1}$ image scale.
The CCD was used with a gain corresponding to $\rm {3.8}$ electrons per
CCD adu, and the readout noise estimated from the
variance of intensity values in the overscan region was measured to be
$\rm \sim {8.1}$ electrons. The CCD saturates at $\rm \sim {30000}$ electrons.
Seeing was measured to be $\sim 2''$ FWHM typically.
Bias frames were taken at the beginning
and end of each observing night and flat field exposures of the dome
interior were also obtained.

\section{Data Analysis}

The analysis process primarily fell in three parts: firstly, we performed
basic photometric calibration of the data,
secondly we made high signal--to--noise
images suitable to locate any interesting patches at the site of
(or close to) the supernova, thirdly
we performed photometric measurements of features visible
in the SN environments.

\subsection{Calibration}

First, following standard procedures, the images were
debiassed and flat--fielded (using IRAF arithmetic routines). Then
spatial registration of the images was performed.
For each galaxy, one image was taken as reference.
Then the IRAF task IMEXAM was used to measure the pixel coordinates
of the peak intensity of an object (or objects)
visible in all images (and for all bands), using a centering algorithm.
If the nucleus of the galaxy was not suitable, i.e. it was saturated,
then the positions of typically three stars were located and taken as
reference,
otherwise the position of the galaxy nucleus itself was used.
All images of each galaxy were then registered with linear shifts and
linear interpolation of fluxes (IMSHIFT or IMLINTRAN in IRAF).

After the registration, the sum of the images in the same band was derived to
maximize the signal to noise ratio of each color.
For each galaxy one B sum, one V sum and one R sum were produced.
In addition, we also simply summed all available data in all bands
to get a maximum $\rm S/N$ intensity image.

The sky levels were estimated
with the use of the ``median'' option of IMSTAT (in IRAF) in an empty window
of these images selected by eye.
The photometric measurements were made by using PHOT (in noao.digiphot.apphot).
PHOT was used for two different operations: 
to use the CCD image to simulate aperture photometry of the entire galaxy
and hence, by comparison to published photometry, derive an absolute
calibration of the data; and secondly
to obtain aperture photometry of all patches visible around the location of
the supernova to investigate the SN environment and locate candidate light
echoes.
The photometry used for absolute calibration
 was from the Longo \& de Vaucouleurs (1983) and the de
Vaucouleurs \& Longo (1988) aperture photometry catalogs (respectively B and
V, and R), not corrected for galactic extinction. The photometric
calibration in each band for each individual galaxy was derived by taking
the mean value of all apertures in that band.
Magnitudes indicated as ``uncertain'' in the catalogs were not used.

Five galaxies (NGC 3115, 3627, 4303, 4382 and 6384) were saturated
in their nuclear regions in the long exposures.
Thus the calibration for these was bootstrapped from a single unsaturated
short exposure, which was calibrated as above.

No aperture photometry was found in the literature for
UGC 2069, UGC 2259, IC 4237, IC 4798 and NGC 4674.
The calibration was realized by adopting the total-extinction corrected
blue magnitude provided by NED (the NASA Electronic Database)
and by adopting standard
B-V and V-R colors appropriate to the galaxy morphological types.
Aperture magnitudes were derived for aperture values appropriate
to each run pixel scale.

The calibration for NGC 1058 and NGC 1073 was done with respect to two
calibration stars (one observed in each night
of observations). In the case of NGC 1058 this was necessary because
literature photometric apertures fell partially outside the region imaged.

The calibration for field A of NGC 6946 was obtained by scaling it
to the calibration of field B, because the galaxy nucleus
does not appear in field A, although the two fields imaged do overlap.

Finally for a number of galaxies, indicated with an asterisk in Tables 1a and
1b, R magnitudes were not available. In these cases we assumed a V-R
appropriate
to the morphological type of the galaxy, using the relations given
by Buta \& Williams (1995; their Table 6 and Figure 3). 

\subsection{Optical structure}

To locate patches of faint light in order to
investigate the environments of the supernovae and seek candidate echoes,
we firstly simply summed all images in all bands of a given field to make
an image with essentially maximum
signal to noise (S/N): the ``total sum'' image.
We filtered the image with a 3 X 3 median to eliminate
cosmic rays. To isolate compact, discrete patches of emission from the
underlying larger scale complex emission of the host galaxy, we used
a digital ``unsharp masking'' method. The high S/N galaxy image
was differenced with a 17 X 17 median filtered version of the same image.
The final output is an image well suited to identify patches of
faint light against the background. 

In addition to unsharp mask images of the high S/N total sum, we also derived
in the same way, unsharp mask images of each individual bandpass
(in B,V and R). Figures 1 through 36
show the environments of the individual supernovae as derived
from these unsharp masked images: Figs. 1 through 13 are for the JKT
observations, the remaining for the ESO. The figures are in order of
the galaxy Right Ascension.
Each supernova environment is shown in 4 different images:
the total sum, followed by the three bands, B, V, and  R.
A 30 arcsec side box is shown centered at the
SN position and displayed between ${-2}{\sigma}$ and ${+3}{\sigma}$
where ${\sigma}$ is a robust estimate of the dispersion of intensity
levels in the box (specifically half the percentile width corresponding
to $\pm {1 {\sigma}}$ for a Gaussian distribution).

\subsection{The supernova environments}

Each of the unsharp masked total sum images was visually inspected
and the position of all distinct objects or patches within a
10 arcsec diameter circle centered at the supernova were tabulated.
In a few instances the position of patches located outside of the ring
was tabulated when these patches appear blue in the images.
{\it Keeping these tabulated positions fixed}, we then measured
their brightness in the other three B, V and R images using a fixed
aperture of 3 pixel radius for the JKT run and a fixed aperture of
2 pixels for the ESO run (corresponding to $0.99$ and $1.22$ arcsec
respectively).

The limiting magnitude for each box is different due to
different photometric conditions, different telescopes,
exposure times, and different levels of contamination
from the underlying host galaxy. Approximate B, V and R limiting magnitudes are
respectively given in the last three columns of Tables 2a and 2b.
They correspond to a ${\rm 5{\sigma}}$ detection, unless otherwise
indicated.
Each limiting magnitude is calculated over an aperture of 3 and 2
pixels in radius respectively for the JKT and the ESO runs
using the ${\sigma}$ which is the robust estimate of the dispersion of
intensity levels in the box (as discussed above).
A complete discussion of Tables 2a and 2b is given in Section 4.

Local Galactic extinction corrections were also applied to the resulting
photometry, using values
taken from the $\rm A_B$ of the Third Reference Catalog (RC3;
de Vaucouleurs et al. 1986).
The corresponding V and R corrections were derived by taking the
following relationships from Cardelli et al. (1989):
${\rm {A_B} = {1.324}\times{A_V}}$ and ${\rm {A_R} = {0.84}\times{A_V}}$.


\section{Results}

\subsection{Basic data}

Results from the photometric data reduction are summarized in Tables 2a and 2b
(for the JKT and the ESO runs respectively).
In the first column the name of the galaxy is given; then the supernova(e)
and the SN Type are listed in columns two and three.
In the fourth column we list all patches that have been
identified at (or close to) the site of the supernova and that
were bright enough to be detected during the observation: they are numbered
and briefly described by a comment that follows in column five.
Such features are close to/within a ring of approximately 5 arcsec in
radius centered at the nominal position of the supernova (see Figs.
1 through 36).
A major uncertainty in this work is the actual location
of the historical supernova. Re--measurement of a small number of plates
containing images of the actual event (ongoing work) suggests
typical uncertainties of the order of a few arcsec. Hence, for the present
purposes, we chose to study a region of $\rm {10} ~arcsec$ diameter
centered at the nominal location of the SN.
For each patch the offset coordinates (in arcsec) with respect to the SN
position have been calculated and are given in the next two columns.
The estimated B, V and R magnitudes and the B-V and V-R colors are listed in 
columns 8 through 12 of these Tables. They are all corrected for local Galactic
extinction.
As mentioned in Sect $3.3$, the following three columns give respectively the
B, V and R limiting magnitudes corresponding to a ${\rm 5 {\sigma}}$ detection
(but some magnitudes correspond to a ${\rm 3 {\sigma}}$ detection limit
and are indicated with an asterisk in Tabs. 2a and 2b).
Finally the last three columns give the estimated ${\rm 1 {\sigma}}$
uncertainties of
the measured B, V and R magnitudes.
These uncertainties lead to a typical uncertainty of $0.2$ magnitude
in the colors $\rm B-V$ and $\rm V-R$ (see color/color plots to follow).

The positions of a handful of SNe are accurately known from radio observations
or astrometric measurements.
Pennington et al. (1982) provide accurate astrometric
positions of SNe 1923A and 1957D (in NGC 5236).
For other two supernovae in the same galaxy, SNe 1950B and 1983N, accurate
positions are derived from radio observations (Wei\-ler et al. 1986).
Given these absolute positions, the supernovae were located with respect
to the SN 1957D position, which was identified by looking at the SN field
observed by Long et al. (1992).
The position of SN 1960L is accurately known from astrometry
(Porter 1993). In the case of SN 1989B we have collected a wide variety
of positional information. The position of SN 1961V in NGC1058
well agrees with the radio position given in Cowan et al. (1988).
Finally the position of SN 1885A in NGC 224 (M31) is in very good
agreement with the coordinates of its supernova remnant as measured
from Hubble Space Telescope Wide Field Planetary Camera-2
images by Fesen et al. (1998). Hence, in these cases, we are much more
confident of the location of the original event in our images.


\subsection{Description of Supernova Environments:}

In the following we describe the SN environments presented in Figs. 1
through 36. All sites are carefully analyzed and plausible candidates
indicated. The 5 arcsec radius ring 
centered at the nominal position of the supernova is depicted at each site.
Each figure can be considered as a finding chart corresponding to each
supernova.
We also include some comparison to
the same environments as observed with the Wide Field Planetary Camera-2
(WFPC-2) in archival Hubble Space Telescope (HST) proposals when available.

{\bf JKT Run:}

{\bf NGC 23 - SN 1955C:} patch 2, south--west of the SN, is red in B-V
and is within imaging limits
from the ground. It is probably just part of the body of the galaxy.

{\bf NGC 210 - SN 1954R:} one patch is clearly visible at the northern edge of
the ring; it is very blue in both colors.
Classified as a candidate (see Table 3).

{\bf NGC 224 - SN 1885A:} five small patches ($\le 1$ arcsec)
have been detected. They all are red
and bright (brighter than 20th V magnitude).
The galaxy light might contaminate the environment.

{\bf NGC 253 - SN 1940E:} two bright patches; in both cases the B-V is very red
and the V-R very blue. Patch 1 appears compact and
isolated; patch 2 is extended and arc--like. More structure is visible
outside of the 5 arcsec ring and is possibly just part of the arm of the
galaxy.

{\bf NGC 488 - SN 1976G:} one very red and bright extended patch (far off
North to the assumed SN position). It looks like a small tail ``departing''
from the galaxy nucleus.

{\bf UGC 2069 - SN 1961P:} no patches are distinguishable within the
region just around the SN;
an isolated compact object (possibly a star?) is to the south-east
of the SN nominal position.
An arc of compact bright objects lies well to the northern edge
of the SN sections.

{\bf UGC 2105 - SN 1938A:} two patches are located, one to the south-west
of the supernova (within the $\rm 5 ~arcsec$ ring and the other well
further out in the same direction. The latter looks very blue from visual
inspection. The photometry indicates that the first patch is
blue in color, and that the second one is very blue and faint
in visual magnitude.
They are both brighter than ${\rm 3 {\sigma}}$ detection limits.

{\bf NGC 1003 - SN 1937D:} two patches are identified
in the immediate surroundings of the SN location: they are both displaced
in the x coordinate from the SN position and are compact. The hosting galaxy
was oberved with WFPC-2 on the HST. The region around the SN is not very
crowded and compact sources of emission are visible. Two groups
of stars are observed South of the SN immediate environment
and appear unrelated to the region of interest.

{\bf NGC 1058 - SNe 1961V, 1969L:} an elongated patch is clearly visible
{\it at} the location of SN 1961V; in V it appears broken up into different
smaller patches: a  couple of these are faint and extremely blue in B-V.
The region of SN 1969L is empty.
SN 1961V classified as a candidate (see Table 3).

{\bf NGC 1073 - SN 1962L:} no patches are visible visible
in the immediate surroundings of the SN.
An extended patch instead is visible off the $\rm 5 ~arcsec$ ring
to the south--east of the SN position.

{\bf UGC 2259 - SN 1963L:} a blue and compact object
is visible to the south--east of the SN position on the edge
of the ring. Two patches slightly off the ring, to the north--west,
appear blue and patchy and should be further investigated.
Classified as a candidate (see Table 3).

{\bf NGC 1325 - SN 1975S:} no patches are visible either within or
outside the ring.

{\bf NGC 2276 - SNe 1962Q, 1968V, 1968W:} no R band observations are available.
The first two SNe lie in crowded regions. South--east of SN 1962Q a
bright and compact patch of light is seen. At least four relatively bright
patches are identified at around the location of SN 1968V. Further
observations are needed (in the R band for example).
At the SN 1968W position some blue emission brighter than 
a ${\rm 3 {\sigma}}$ magnitude is visible.
The last supernova is close to the nucleus of the galaxy.

{\bf NGC 6946/field A+field B - SNe 1917A, 1939C, 1948B, 1968D, 1969P, 1980K:}
at the sites of SNe 1948B, 1968D and 1980K
some patches are located: they all are bright and quite blue.
SN 1980K is likely the most
interesting of all SNe in this galaxy; in fact from all the data collected
at radio wavelengths over a long period
of time (Weiler et al. 1986, 1991) and from the fact that its optical
counterpart has also been identified, it appears a good candidate to look
at because of dense circumstellar and interstellar environments.
At the sites of SNe 1917A and 1969P bright blue patches are observed.
Star--like objects appear in the environment. In the case of SN 1939C
some objects are visible as well.
SNe 1917A, 1969P and 1980K are classified as candidates (see Table 3).

{\bf NGC 7177 - SNe 1960L, 1976E:}
at the location of SN 1960L some patches are seen. A blue one is seen {\it at}
the location of the supernova, brighter than a ${\rm 5 {\sigma}}$ detection.
As already discussed the position of SN 1960L is accurately known from
Porter (1993) and thus the identification of a blue patch at the location
of the supernova is of great interest and calls for further investigation.
At the site of SN 1976E a very bright and red compact object is observed. 
SN 1960L is classified as a candidate (see Table 3).

{\bf NGC 7331 - SN 1959D:} patch 1 is within a  ${\rm 3 {\sigma}}$
detection limit; patch 2 is brighter and blue, although appears
compact on the image and might be only a star. Patch 3 is also blue and
compact. These patches are also visible on the archival WFPC-2 image
that we analyzed. With WFPC-2 it is clear that there are groups of
stars present (possibly new forming stars) and/or HII regions belonging to
a galaxy spiral arm.

{\bf ESO Runs:}

{\bf NGC 2935 - SN 1975F:} one extremely bright compact object not too far
off the SN position; red in B-V. It looks similar to what is seen at the
location of SN 1976E in NGC 7177.

{\bf NGC 3115 - SN 1935B:} one red and compact object is visible to the
south--east of the SN location and at the edge of the ring. Several other
patches of the same kind are found all around the region.

{\bf NGC 3627 - SNe 1973R, 1989B:} in the case of SN 1973R a bright blue
elongated object is visible. Almost {\it at} the location of SN 1989B
a roundish fuzzy and very blue (in both B-V and V-R) patch of light is
present. The emission we see is within a ${\rm 3 {\sigma}}$ detection limit.
Both SN environments are part of spiral arms and very crowded.
On the basis of our selection criteria,
both supernovae are classified as candidates (see Table 3).

{\bf NGC 4038 - SNe 1921A, 1974E:} no patches are visible within 5 arcsec
of the nominal positions of these two supernovae.

{\bf NGC 4254 - SNe 1967H, 1972Q, 1986I:} several bright and blue
features are seen at the locations of these supernovae. The SN sites are
quite crowded though
and the visible features might simply belong to the spiral arms of the galaxy.

{\bf NGC 4303 - SNe 1926A, 1961I, 1964F:} at all these sites some
features are present. From visual inspection and its blue color,
the faint arc--shaped patch at the SN 1926A site looks promising.
It is brighter than a ${\rm 3 {\sigma}}$ detection limit.
The extended, bright, round
patch at the site of SN 1961I is quite likely to be a HII region or young open
cluster. SN 1926A is classified as a candidate (see Table 3).
Both SNe 1926A and 1964F are visible in the WFPC-2 images of the host galaxy.
We believe we can confirm what we noticed in our ground based
observations.

{\bf NGC 4321 (M100) - SNe 1901B, 1914A, 1959E, 1979C:} all environments
show relatively bright and blue patches of light. The SN 1979C
environment is possibly the most promising, as it is known from radio
observations to be characterized by a dense environment
(Weiler et al. 1981, 1986). This SN has also been optically identified.
Only in this one case (in this galaxy) is an emission patch clearly visible
within 5 arcsec from the SN nominal position.
Less compelling evidence has been gathered for the other supernovae.
SN 1979C is classified as a candidate (see Table 3).
The positions of SNe 1959E and 1979C are also identified on some archival
WFPC-2 observations. Multiple sources are present in the immediate
surroundings of the SNe. Many may well be star clusters,
others more directly related to the SN event.
For example we know the optical counterpart to
SN 1979C has been identified among all other sources (Van Dyk, private
communication).

{\bf NGC 4382 - SN 1960R:} two very faint patches are detected within the
immediate SN regions. All images appear to be
very noisy. Patch 1 is within a ${\rm 3 {\sigma}}$ detection limit.

{\bf NGC 4424 - SN 1895A:} from both runs (March and May 1995)
one faint and compact patch
is visible to the south--east of the SN location (at the edge of the
$\rm 5 ~arcsec$ ring); it is very red in B-V.

{\bf NGC 4674 - SN 1907A:} an elongated structure
lying to the south--west of the SN is visible.
Some other patch is also visible far off the SN nominal position.
Both are red in color and possibly are just part of the main
body of the galaxy.

{\bf NGC 4753 - SN 1965I, 1983G:} at both locations some structure is
found. The March 1995 run is very noisy and does not provide any information.
Only at the location of SN 1965I (May run) is some faint structure visible
to the south--west of the supernova.

{\bf IC 4237 - SN 1962H:} a bright, blue, roundish patch is visible in all
sections and is located (within 5 arcsec from the SN position)
at the end of an arc--like structure
possibly part of a spiral arm of the galaxy.
Classified as a candidate (see Table 3).

{\bf NGC5236 (M83) - SNe 1923A, 1950B, 1957D, 1968L, 1983N:}
all patches at all locations appear to be very bright and blue (both
in the B-V and in the V-R colors).
{\it At} the site of SN 1983N two compact
patches are visible.
Note that the positions of these supernovae are known accurately
(see Section $4.1$).
In all these cases the SN star progenitors are likely to have been
massive (two SNe are confirmed Type II's and one is a
Type Ib; SN 1957D is likely to be a Type II, as possibly inferred from the
radio data, see Weiler et al. 1986). Around such progenitors dense
circumstellar and possibly interstellar environments are expected,
thus allowing for the formation of light echoes.
The best bets may be SNe 1957D and 1983N.
SNe 1957D and 1983N are classified as candidates (see Table 3).

{\bf NGC 5253 - SN 1895B, 1972E:} no R band observations are
available.  Some structure is present close to
the location of SN 1895B. Some patches are also visible near the location
of SN 1972E. SN 1895B is also on the edge of some WFPC-2 images.
As single images in different filters were found the image quality
of the summed image is not optimal. We can see some structure
in the surroundings of the supernova but this HST imaging does not contribute
additional information on the presence of reflected light.

{\bf NGC 5668 - SNe 1952G, 1954B:} both show some faint structures slightly
north of the SN positions. 
The surrounding environments are crowded.
Both supernovae lie not too far from the nucleus of the galaxy and belong to
spiral arms. The May observations are slightly affected by saturation;
the March ones have a poorer signal--to--noise ratio
(the galaxy was observed well into twilight).
For the March run no R band observations are available.
WFPC-2 observations resolve the underlying spiral arm structure.
The environments are not particularly crowded and at the location
of SN 1952G the same single bright object observed from the ground is also
observed with HST, i.e. it remains unresolved with WFPC-2.

{\bf NGC 5857 - SN 1950H, 1955M:} a compact and blue patch
is found at the location of SN 1950H.
This SN lies close to the nuclear part of the galaxy.

{\bf NGC 5861 - SN 1971D:} a patch is visible within the $\rm 5 ~arcsec$
ring, very close
to the location of the SN. Classified as a candidate (see Table 3).
WFPC-2 observations show emission close to the SN position.
Yet the HST images resolve the spiral arm and thus many
compact objects are indeed visible, making it difficult to
discriminate easily among them.

{\bf NGC 6181 - SN 1926B:} a blue and bright double--lobed object is visible
at the site of this supernova. It lies within the $\rm 5 ~arcsec$
ring and occupies
more than half of it. Classified as a candidate (see Table 3).
This SN lies on the top left corner of the PC images
retrieved from the archive. The SN environment is not very crowded,
at the extreme end of the spiral arm. Some faint compact sources of
emission are visible.

{\bf NGC 6384 - SN 1971L:} a blue arc--shaped patch
is present just close to the center of the section and is brighter than the
${\rm 3 {\sigma}}$ detection limit. HST images resolve the underlying
spiral structure and several compact sources of emission are present
within the uncertainty ring. Some structure is seen around the
nominal position of the SN.

{\bf IC 4719 - SN 1934A:} along the circumference of the $\rm 5 ~arcsec$
ring a series of faint
and blue knots are visible; some are within  a ${\rm 5 {\sigma}}$
detection and some within a ${\rm 3 {\sigma}}$.

{\bf IC 4798 - SN 1971R:} two bright patches are visible, one, patch 1,
possibly belongs to the nuclear part of the galaxy. The second one is
to the south--west of the SN.

{\bf NGC 6835 - SN 1962J:} a patch is visible within the ring, to the east
of the supernova. It is 
within a ${\rm 3 {\sigma}}$ detection limit.

{\bf NGC 7177 - SNe 1960L, 1976E:}
at the location of SN 1960L some patches are  seen as in the JKT observations.
A blue one is seen at the
location of the supernova and is within a
${\rm 3 {\sigma}}$ detection limit.
SN 1960L classified as a candidate (see Table 3).

\subsection{Candidate Light Echoes:}

On the basis of what we see
at various SN positions, we classify some optical emission patches as likely
candidate light echoes and present them in Table 3.
These candidates are close to the SN nominal position (typically within
the 5 arcsec ring, but some lie as far as 6 arcsec away),
appear blue in color and are compact. A bonus is granted when the
SN position is
accurately known (see section ${4.1}$).
All data in Tab. 3 are derived from Tables 1a through 3b.


\subsection{Correlations and SN environments:}

In order to investigate the SN environments we have made color/color
plots for the features visible at the SN environment.
In Fig. 37
we plot the two color diagram of all patches  whose
B, V and R magnitudes are brighter than the
calculated limiting magnitudes (corresponding either to a ${\rm 5 {\sigma}}$
or a ${\rm 3 {\sigma}}$ detection as presented in Sections $3.3$ and $4.1$).
The ESO observations are the open circles (${\rm 5 {\sigma}}$;
40 objects) and the open squares (${\rm 3 {\sigma}}$; 11 objects);
the JKT are the filled circles (${\rm 5 {\sigma}}$; 37 objects)
and the filled squares (${\rm 3 {\sigma}}$; 2 objects);
SN 1991T (data from Schmidt et al. 1994)
is represented by the filled triangles.
All patches appear distributed within the following intervals:
$\rm {-1.2} < (B-V) < {1.8}$ and  $\rm {-0.8} < (V-R) < {1.6}$,
centered at around ${0.3}$ in both colors. Typically both colors are
affected by a $0.2$ magnitude error as can be seen in Fig. 38,
where for each data point the error bars estimated from the ${\rm 1 {\sigma}}$
uncertainties in Tabs. 2a and 2b are plotted.



Figure 39 is the same as Fig. 37 but allowing for a
differentiation
of the SN Types: all filled symbols refer to Type I SNe
(Ia's), all open ones to Type II's (Type Ib's are inserted in this
group as they likely derive from similarly massive star progenitors;
in our samples there are no Type Ic SNe).
The ESO observations are the squares; the JKT, the circles.
The empty triangles refer to 
peculiar supernovae and those whose Type is unknown according  to the
classification in Tables 2a and 2b.
The filled triangles are SN 1991T (from Schmidt et al. 1994).
We have always followed the spectroscopic classification
of Branch (1990) and therefore
SNe 1961I, 1961V and 1964F
have been considered Type II, while SNe
1940E, 1960L, 1963L and 1967H unknown.
SN 1961V (in NGC 1058) which was classified as a Type V by Zwicky (1965)
might not be a pure SN event (Goodrich et al. 1989, Filippenko et al. 1995)
and yet can be as useful as any other object if it produced a
light echo.
It is clear that different Types of
SNe are located in different regions of the color/color plot:
the emission patches associated to
Type Is are mostly located at B-V and V-R redder than $\sim{0.10}$,
while Type IIs are on average bluer than this value in
both colors. Unknown/peculiars are randomly distributed;
the location of the peculiar Type Ia SN 1991T is somewhat consistent
with the other peculiars.
This is what might be expected from the distribution of different Type SNe in
different galaxies and in different parts of the parent galaxies themselves.
SNe II/Ib likely derive from massive progenitors in spiral galaxies
and thus are associated to star--formation regions where a strong
H$\alpha$ emission is present (Van Dyk et al. 1996 and references therein).
This would tend to cause a high positive value of $\rm (V-R)$.
Type Ia SNe, instead, possibly derive
from a wider variety of stellar environments (as they are observed both in
elliptical and spiral galaxies; see \cite{B} for references).


In Fig. 40
we plot all supernovae except the un\-known/pe\-cu\-liar.
We note that known Type SNe cover a much smaller portion of the
color/color plane than unknown/peculiar.


In Fig. 41 we plot for comparison both a
reddening vector (solid arrow)
and a scattering vector (dotted arrow) to ascertain in what proportions
these two physical components affect the colors of our sample of patches.
The reddening vector was derived by assuming the Rieke \& Lebofski (1985)
reddening law: $\rm E(B-V)/E(V-R)={0.78}$.
The effect of scattering, also, was estimated using scattering cross--sections
as tabulated in Sparks (1994). With these numbers and assuming
wavelengths of 4400, 5500, 6400 for the BVR filters, light which has
been scattered is always bluer by $\rm {-0.3}$ in B-V and $\rm {-0.2}$
in V-R than the unscattered illuminating source.
For illustration, the scattering vector shown originates
at B-V=V-R=0, which is a reasonable estimate of the colors of
a supernova at peak luminosity in many cases
(D. Branch private communication). In fact, we cannot be sure in detail
what the colors of the time integrated supernova light curves are since decay
rates can be quite wavelength dependent, and reddening may also play
a role. Nevertheless, outliers, and especially blue outliers,
in these diagrams should be considered as objects worthy of additional
follow--up observations as potential echoes,
and yet strict SN 1991T look-alikes, i.e. definite echoes (\cite{Setal}),
are not present in the sample.


Finally, for reference, in Fig. 42
colors of Main Sequence stars of
all spectral types
were plotted over our sample (these colors were taken from \cite{Z92}).
The distribution of patches is broadly
consistent with the locations of these Main Sequence stars.
This distribution covers also a similar region to a composite
stellar population as a function of age as was found by
plotting an evolutionary track kindly provided by C. Leitherer.
We therefore infer that for the most part we are simply seeing stars or
star clusters. However there are in addition patches with both
unusually red and unusually blue colors which should be
considered with special care.


\section{Conclusions and Future Work}

We have looked for candidate light echoes around 64 supernovae of
all Types in 38 galaxies. From this ground--based photometric data we notice
optical emission at the SN nominal positions.
The color distribution of these patch\-es is broad,
and generally consistent with stellar population colors, possibly with
some reddening. However there are in addition patches with both
unusually red and unusually blue colors.  We expect light echoes to be
blue, and while none of the objects are quite as blue in V-R as the
known light echo of SN1991T, there are features that are unusually blue
and we identify these as candidate echoes for follow-on space
and ground observations.

We also made reference to archival WFPC-2/HST observations 
of fourteen SN fields.
These showed that some of the patches resolve into multiple point sources,
while others do not. The absence of multi-color information within the
HST data precludes us from narrowing down our candidate list further,
since at WFC resolution we do not expect the SN remnants to show significant
spatial extent. As discussed in the text and in Sparks (1994) only
polarimetric observations of a field many arcsec in extent around the SN
position can succeed in determining whether (or not) there is an
extended structure. This is the basis for future observations using the
Advanced Camera on the Hubble Space Telescope.

From the color/color plots we notice that indeed different SN Types
are characterized by different environments.
SNe Ia tend to be redder than $\rm (B-V)=0-{0.1}$ and $\rm (V-R)=0-{0.1}$,
while Type Ib/II are bluer than these values.
This distribution seems to reflect the distribution of various SN Types
in different regions of the parent galaxies and in different
galaxies as well.
SNe II/Ib likely derive from massive progenitors in spiral galaxies
and thus are associated to star--formation regions where a strong
H$\alpha$ emission is present (\cite{VD} and references therein).
Type Ia SNe, instead, possibly derive
from a wider variety of stellar environments (as they are observed both in
elliptical and spiral galaxies).



\vspace{1.truecm}

\begin{acknowledgements}
F.R.B. would like to thank
Angela Bragaglia, David Branch, Enrico Cappellaro, Massimo Della Valle,
Boris Dirsch, Claus Leitherer, Arturo Manchado, Ulisse Munari and Elena Pian
for useful discussions and encouragement.
\end{acknowledgements}

\vfill
\eject
\centerline{\bf REFERENCES} 

\begin{list}{}{\leftmargin 20pt} 

\bibitem[Barth et al.\ 1996]{B} Barth A.J., Van Dyk S.D., Filippenko A.V.,
Leibundgut B., Richmond M.W., 1996, AJ 111, 2047

\bibitem[Boffi \& Branch 1995]{BB} Boffi F.R., Branch D., 1995, PASP 107, 347

\bibitem[Branch 1990]{Br90} Branch D., 1990, in
 ``Supernovae'', ed. Petschek A. G., Springer--Verlag, pp. 39--42

\bibitem[Branch et al.\ 1995]{BLYBB} Branch D., Livio M., Yungelson L.R.,
Boffi F.R., Baron E., 1995, PASP 107, 1019
 
\bibitem[Buta \& Williams 1989]{BW} Buta R.J., Williams K.L., 1995, AJ 109, 543

\bibitem[Cardelli et al.\ 1989]{CCM} Cardelli J. A., Clayton G. C.,
Mathis J. S., 1989, ApJ, 345, 245

\bibitem[Chevalier 1986]{Ch} Chevalier R.A., 1986, ApJ 308, 225

\bibitem[Cowan et al.\ 1988]{Cow} Cowan J.J., Henry R.B.C., Branch D., 1988,
ApJ 329, 116

\bibitem[Crotts 1988]{Cr} Crotts A.P.S., 1988, ApJ 333, L51

\bibitem[Fesen et al.\ 1998]{Fe98} Fesen R.A., Gerardy C.L., McLin K.M.,
Hamilton A.J.S., 1998, astro-ph9810002

\bibitem[Filippenko et al.\ 1995]{Fili} Filippenko A.V., Barth A.J.,
Bower G.C., Ho L.C., Stringfellow G.S., Goodrich R.W., Porter A.C., 1995,
AJ 110, 2261

\bibitem[Goodrich et al.\ 1989]{Getal} Goodrich R.W., Stringfellow G.S.,
Penrod G.D., Filippenko A.V., 1989, ApJ 342, 908

\bibitem[Lira et al.\ 1998]{Lira} Lira P., et al. 1998, AJ 115, 234

\bibitem[Long et al.\ 1989]{Long} Long K.S., Blair W.P., Krzeminski W., 1989,
ApJ 340, L25

\bibitem[Longo \& de Vaucouleurs 1983]{Lo83} Longo G., de Vaucouleurs A.,
1983, The University of Texas Monographs in Astronomy no. 3 

\bibitem[Pennington \& Dufour 1983]{PD} Pennington R.L., Dufour R.J., 1983,
ApJ 270, L7

\bibitem[Pennington et al.\ 1982]{PTD} Pennington R.L., Talbot R.J.,
Dufour R.J., 1982, AJ 87, 1538

\bibitem[Perlmutter et al.\ 1998]{Perl} Perlmutter S., et al., 1998,
in preparation

\bibitem[Porter 1993]{Port} Porter A.C., 1993, PASP 105, 1250

\bibitem[Riess et al.\ 1998]{Ri98} Riess A.G., et al., 1998, AJ, 116, 1009
 
\bibitem[Schmidt et al.\ 1994]{Schetal}
Schmidt B.P., Kirshner R.P., Leibundgut B.,
Wells L.A., Porter A.C., Ruiz--Lapuente P., Challis P.,
Filippenko A.V., 1994, ApJ 434, L19 

\bibitem[Rieke \& Lebofski 1985]{RL}
Rieke G.H., Lebofski M.J., 1985, ApJ 288, 618

\bibitem[Sparks 1994]{S94} Sparks W.B., 1994, ApJ 433, 19 (PAPER I)

\bibitem[Sparks 1996]{S96} Sparks W.B., 1996, ApJ 470, 195 (PAPER II)

                     347, L65
\bibitem[Sparks et al.\ 1989]{SPM}
Sparks W.B., Paresce F., Macchetto F., 1989, ApJ 347, L65

\bibitem[Sparks et al.\ 1999]{Setal}
Sparks W.B., Macchetto F., Panagia N., Boffi F.R.,
Branch D., Hazen M.L., Della Valle M., 1999, ApJ, accepted

                    AJ 111, 2017
\bibitem[Van Dyk et al.\ 1996]{VD}
Van Dyk S.D., Hamuy M., Filippenko A.V., 1996, AJ 111, 2017

\bibitem[de Vaucouleurs et al.\ 1986]{deVau86}
de Vaucouleurs  G., de Vaucouleurs A., Corwin H.G.,
Buta R.J., Paturel G., Fouque' P., 1986, `Third Reference Catalog'
(RC3)

\bibitem[de Vaucouleurs \& Longo 1988]{deVau88}
de Vaucouleurs A., Longo G., 1988, The University of
Texas Monographs in Astronomy no. 5

\bibitem[Weiler et al.\ 1981]{Wetal81}
Weiler K.W., van der Hulst J.M., Sramek R.A.,
Panagia N., 1981, ApJ 243, L151 

\bibitem[Weiler et al.\ 1986]{Wetal86}
Weiler K.W., Sramek R.A., Panagia N., van der Hulst 
J.M., Salvati M., 1986, ApJ 301, 790

\bibitem[Weiler et al.\ 1991]{Wetal91}
Weiler K.W., Van Dyk S.D., Panagia N., Sramek R.A.,
Discenna J.L.,  1991, ApJ 380, 161

\bibitem[Zombek 1992]{Z92}
Zombek M.V., 1992, Handbook of Space Astronomy and
Astrophysics, Cambridge University Press, Chapter 2, p. 68

\bibitem[Zwicky 1965]{Zw} Zwicky F., 1965, in ``Stars and Stellar Systems'',
Vol. VIII (Stellar Structure), eds. L. H. Aller and D. B. McLaughlin
(Chicago: University of Chicago Press), p. 367

\eject

\end{list}

\vfill
\eject

\centerline{\bf FIGURE CAPTIONS}

\medskip

{\bf Fig. 1}:
 TOP: NGC 23 and the environments of SN 1955C.
BOTTOM: NGC 210 and the environments of SN 1954R. Each environment is taken as
a 30 arcsec side box centered at the SN position. The ring is 5 arcsec
in radius. See text for details.

\medskip

{\bf Fig. 2}:
 TOP: NGC 224 and the environments of SN 1885A.
BOTTOM: NGC 253 and the environments of SN 1940E. Each environment is taken as
a 30 arcsec side box centered at the SN position. The ring is 5 arcsec
in radius. See text for details.

\medskip

{\bf Fig. 3}:
 TOP: NGC 488 and the environments of SN 1978G.
BOTTOM: UGC 2069 and the environments of SN 1961P. Each environment is taken as
a 30 arcsec side box centered at the SN position. The ring is 5 arcsec
in radius. See text for details.

\medskip

{\bf Fig. 4}:
 TOP: UGC 2105 and the environments of SN 1938A.
BOTTOM: NGC 1003 and the environments of SN 1937D. Each environment is taken as
a 30 arcsec side box centered at the SN position. The ring is 5 arcsec
in radius. See text for details.

\medskip

{\bf Fig. 5}:
 TOP: NGC 1058 and the environments of SN 1961V.
BOTTOM: NGC 1058 and the environments of SN 1969L. Each environment is taken as
a 30 arcsec side box centered at the SN position. The ring is 5 arcsec
in radius. See text for details.

\medskip

{\bf Fig. 6}:
 TOP: NGC 1073 and the environments of SN 1962L.
BOTTOM: UGC 2259 and the environments of SN 1963L. Each environment is taken as
a 30 arcsec side box centered at the SN position. The ring is 5 arcsec
in radius. See text for details.

\medskip

{\bf Fig. 7}:
 TOP: NGC 1325 and the environments of SN 1975S.
BOTTOM: NGC 2276 and the environments of SN 1962Q. Each environment is taken as
a 30 arcsec side box centered at the SN position. The ring is 5 arcsec
in radius. See text for details.

\medskip

{\bf Fig. 8}:
 TOP: NGC 2276 and the environments of SN 1968V.
BOTTOM: NGC 2276 and the environments of SN 1968W. Each environment is taken as
a 30 arcsec side box centered at the SN position. The ring is 5 arcsec
in radius. See text for details.

\medskip

{\bf Fig. 9}:
 TOP: NGC 6946 and the environments of SN 1917A.
BOTTOM: NGC 6946 and the environments of SN 1939C. Each environment is taken as
a 30 arcsec side box centered at the SN position. The ring is 5 arcsec
in radius. See text for details.

\medskip

{\bf Fig. 10}:
 TOP: NGC 6946 and the environments of SN 1948B.
BOTTOM: NGC 6946 and the environments of SN 1968D. Each environment is taken as
a 30 arcsec side box centered at the SN position. The ring is 5 arcsec
in radius. See text for details.

\medskip

{\bf Fig. 11}:
 TOP: NGC 6946 and the environments of SN 1969P.
BOTTOM: NGC 6946 and the environments of SN 1980K. Each environment is taken as
a 30 arcsec side box centered at the SN position. The ring is 5 arcsec
in radius. See text for details.

\medskip

{\bf Fig. 12}:
 TOP: NGC 7177 and the environments of SN 1960L.
BOTTOM: NGC 7177 and the environments of SN 1976E. Each environment is taken as
a 30 arcsec side box centered at the SN position. The ring is 5 arcsec
in radius. See text for details.

\medskip

{\bf Fig. 13}:
 TOP: NGC 7331 and the environments of SN 1959D.
Each environment is taken as
a 30 arcsec side box centered at the SN position. The ring is 5 arcsec
in radius. See text for details.

\medskip

{\bf Fig. 14}:
 TOP: NGC 2935 and the environments of SN 1975F.
BOTTOM: NGC 3115 and the environments of SN 1935B. Each environment is taken as
a 30 arcsec side box centered at the SN position. The ring is 5 arcsec
in radius. See text for details.

\medskip

{\bf Fig. 15}:
 TOP: NGC 3627 and the environments of SN 1973R.
BOTTOM: NGC 3627 and the environments of SN 1989B. Each environment is taken as
a 30 arcsec side box centered at the SN position. The ring is 5 arcsec
in radius. See text for details.

\medskip

{\bf Fig. 16}:
 TOP: NGC 4038 and the environments of SN 1921A.
BOTTOM: NGC 4038 and the environments of SN 1974E. Each environment is taken as
a 30 arcsec side box centered at the SN position. The ring is 5 arcsec
in radius. See text for details.

\medskip

{\bf Fig. 17}:
 TOP: NGC 4254 and the environments of SN 1967H.
BOTTOM: NGC 4254 and the environments of SN 1972Q. Each environment is taken as
a 30 arcsec side box centered at the SN position. The ring is 5 arcsec
in radius. See text for details.

\medskip

{\bf Fig. 18}:
 TOP: NGC 4254 and the environments of SN 1986I.
BOTTOM: NGC 4303 and the environments of SN 1926A. Each environment is taken as
a 30 arcsec side box centered at the SN position. The ring is 5 arcsec
in radius. See text for details.

\medskip

{\bf Fig. 19}:
 TOP: NGC 4303 and the environments of SN 1961I.
BOTTOM: NGC 4303 and the environments of SN 1964F. Each environment is taken as
a 30 arcsec side box centered at the SN position. The ring is 5 arcsec
in radius. See text for details.

\medskip

{\bf Fig. 20}:
 TOP: NGC 4321 (M100) and the environments of SN 1901B.
BOTTOM: NGC 4321 and the environments of SN 1914A.
Each environment is taken as
a 30 arcsec side box centered at the SN position. The ring is 5 arcsec
in radius. See text for details.

\medskip

{\bf Fig. 21}:
 TOP: NGC 4321 and the environments of SN 1959E.
BOTTOM: NGC 4321 and the environments of SN 1979C. Each environment is taken as
a 30 arcsec side box centered at the SN position. The ring is 5 arcsec
in radius. See text for details.

\medskip

{\bf Fig. 22}:
 TOP: NGC 4382 and the environments of SN 1960R.
BOTTOM: NGC 4424 (May) and the environments of SN 1895A.
Each environment is taken as
a 30 arcsec side box centered at the SN position. The ring is 5 arcsec
in radius. See text for details.

\medskip

{\bf Fig. 23}:
 TOP: NGC 4424 (March) and the environments of SN 1895A.
BOTTOM: NGC 4674 and the environments of SN 1907A. Each environment is taken as
a 30 arcsec side box centered at the SN position. The ring is 5 arcsec
in radius. See text for details.

\medskip

{\bf Fig. 24}:
 TOP: NGC 4753 (May) and the environments of SN 1965I.
BOTTOM: NGC 4753 (May) and the environments of SN 1983G.
Each environment is taken as
a 30 arcsec side box centered at the SN position. The ring is 5 arcsec
in radius. See text for details.

\medskip

{\bf Fig. 25}:
 TOP: NGC 4753 (March) and the environments of SN 1965I.
BOTTOM: NGC 4753 (March) and the environments of SN 1983G.
Each environment is taken as
a 30 arcsec side box centered at the SN position. The ring is 5 arcsec
in radius. See text for details.

\medskip

{\bf Fig. 26}:
 TOP: IC 4237 and the environments of SN 1962H.
BOTTOM: NGC 5236 (M83) and the environments of SN 1923A.
Each environment is taken as
a 30 arcsec side box centered at the SN position. The ring is 5 arcsec
in radius. See text for details.

\medskip

{\bf Fig. 27}:
 TOP: NGC 5236 and the environments of SN 1950B.
BOTTOM: NGC 5236 and the environments of SN 1957D. Each environment is taken as
a 30 arcsec side box centered at the SN position. The ring is 5 arcsec
in radius. See text for details.

\medskip

{\bf Fig. 28}:
 TOP: NGC 5236 and the environments of SN 1968L.
BOTTOM: NGC 5236 and the environments of SN 1983N. Each environment is taken as
a 30 arcsec side box centered at the SN position. The ring is 5 arcsec
in radius. See text for details.

\medskip

{\bf Fig. 29}:
 TOP: NGC 5253 and the environments of SN 1895B.
BOTTOM: NGC 5253 and the environments of SN 1972E. Each environment is taken as
a 30 arcsec side box centered at the SN position. The ring is 5 arcsec
in radius. See text for details.

\medskip

{\bf Fig. 30}:
 TOP: NGC 5668 (May) and the environments of SN 1952G.
BOTTOM: NGC 5668 (May) and the environments of SN 1954B.
Each environment is taken as
a 30 arcsec side box centered at the SN position. The ring is 5 arcsec
in radius. See text for details.

\medskip

{\bf Fig. 31}:
 TOP: NGC 5668 (March) and the environments of SN 1952G.
BOTTOM: NGC 5668 (March) and the environments of SN 1954B.
Each environment is taken as
a 30 arcsec side box centered at the SN position. The ring is 5 arcsec
in radius. See text for details.

\medskip

{\bf Fig. 32}:
 TOP: NGC 5857 and the environments of SN 1950H.
BOTTOM: NGC 5857 and the environments of SN 1955M. Each environment is taken as
a 30 arcsec side box centered at the SN position. The ring is 5 arcsec
in radius. See text for details.

\medskip

{\bf Fig. 33}:
 TOP: NGC 5861 and the environments of SN 1971D.
BOTTOM: NGC 6181 and the environments of SN 1916B. Each environment is taken as
a 30 arcsec side box centered at the SN position. The ring is 5 arcsec
in radius. See text for details.

\medskip

{\bf Fig. 34}:
 TOP: NGC 6384 and the environments of SN 1971L.
BOTTOM: IC 4719 and the environments of SN 1934A. Each environment is taken as
a 30 arcsec side box centered at the SN position. The ring is 5 arcsec
in radius. See text for details.

\medskip

{\bf Fig. 35}:
 TOP: IC 4798 and the environments of SN 1971R.
BOTTOM: NGC 6835 and the environments of SN 1962J. Each environment is taken as
a 30 arcsec side box centered at the SN position. The ring is 5 arcsec
in radius. See text for details.

\medskip

{\bf Fig. 36}:
 TOP: NGC 7177 and the environments of SN 1960L.
BOTTOM: NGC 7177 and the environments of SN 1976E. Each environment is taken as
a 30 arcsec side box centered at the SN position. The ring is 5 arcsec
in radius. See text for details.

\medskip

{\bf Fig. 37}:
 Color/color plot for all patches at/around the location of the
historical supernovae for both observing runs.
In this figure only patches whose
B, V and R magnitudes are brighter than the
calculated limiting magnitudes (corresponding either to a ${\rm 5 {\sigma}}$
or a ${\rm 3 {\sigma}}$ detection, as presented in Section $5.1$)
are plotted.
The ESO observations are the open circles (${\rm 5 {\sigma}}$;
40 objects) and the open squares (${\rm 3 {\sigma}}$; 11 objects);
the JKT are the filled circles (${\rm 5 {\sigma}}$; 37 objects)
and the filled squares (${\rm 3 {\sigma}}$; 2 objects);
SN 1991T is represented by the filled triangles (data from Schmidt et al.
1994).

\medskip

{\bf Fig. 38}:
 The same as Fig. 37 but the corresponding B-V and V-R error bars
are plotted. The corresponding standard deviations in the three band
of observations are given in tables 2a and 2b.

\medskip

{\bf Fig. 39}:
 Figure 39 is the same as Fig. 37 but allowing for differentiation
of the SN Types: all filled symbols refer to Type I SNe
(Ia's), all open ones to Type II's (the only Type Ib SN fulfilling
the selection criteria for this plot is SN 1983N and is plotted 
among the Type II as they likely derive from similarly massive star
progenitors; in our samples there are no Type Ic SNe).
The ESO observations are the squares; the JKT the circles.
The empty triangles refer to 
peculiar supernovae and those whose Type is unknown according  to the
classification in Tables 2a and 2b.
The filled triangles are SN 1991T (data from Schmidt et al. 1994).

\medskip

{\bf Fig. 40}:
 The same as Fig. 39 except that unknown/peculiar SNe are not plotted.

\medskip

{\bf Fig. 41}:
 The same as Fig. 39. The reddening vector (solid line),
derived from the standard
reddening law of Rieke \& Lebofski (1985), and the scattering
vector are also plotted. See text for details.

\medskip

{\bf Fig. 42}:
 The same as Fig. 39. Main Sequence star colors are also plotted for
comparison. See text for details.

\vfill
\eject

\vfill
\eject
\newpage

\centerline{\bf TABLE CAPTIONS}

TABLE 1a:

$^a$ All coordinates are from the revised Shapley-Ames catalog,
unless otherwise noted (3RC=Third Reference Catalog);
$^b$ galaxy morphological type from the Asiago Catalog;
$^c$ offset of the supernova in respect to the nucleus of the galaxy
as given in the Asiago Catalog (units: arcsec);
$^d$ recession velocity from the Asiago Catalog (units: kms$^{-1}$);
$^e$ observation date in format year/month/date;
$^f$ number of exposures taken in each band and exposure times in seconds;
$^g$ two different fields were taken in order to observe all supernovae;
$^h$ offset (in arcsec) in respect to a nearby star (Porter 1993).
An asterisk next to the galaxy name indicates that no R magnitudes were
found in the literature. In these cases to calibrate in R
we assumed a V-R appropriate
to the morphological type of the galaxy, using the relations given
by Buta \& Williams (1995; their Table 6 and Figure 3). 

\medskip

TABLE 1b:

$^a$ All coordinates are from the revised Shapley-Ames catalog,
unless otherwise noted (3RC=Third Reference Catalog);
$^b$ galaxy morphological type from the Asiago Catalog;
$^c$ offset of the supernova in respect to the nucleus of the galaxy as
given in the Asiago Catalog (units: arcsec);
$^d$ recession velocity from the Asiago Catalog (units: kms$^{-1}$);
$^e$ observation date in format year/month/date;
$^f$ number of exposures taken in each band and exposure times in seconds;
the letter ``p'' indicates that polarimetry was done;
$^g$ lost one exposure and then repeated;
$^h$ offset (in arcsec) in respect to a nearby star (Porter 1993);
$^i$ astrometric positions of SNe 1923A and 1957D from Pennington et al.
(1982); radio positions of SNe 1950B and 1983N from Weiler et al. (1986).
An asterisk next to the galaxy name indicates that no R magnitudes were
found in the literature. In these cases to calibrate in R
we assumed a V-R appropriate
to the morphological type of the galaxy, using the relations given
by Buta \& Williams (1995; their Table 6 and Figure 3). 

\medskip

TABLE 2a:

$^a$ The SN Type was taken from Branch 1990 and/or from the Sternberg Catalog;
``:'' after a Type indicates that according to
the Sternberg Catalog there is uncertainty;  both Types are classified
spectroscopically, but the Type II are also classified according to the shape
of the light curve (P=plateau; L=linear) as given by the Sternberg Catalog.
This means that whenever a Type II is indicated as a Type IIP or IIL
the Sternberg classification was used;
for some objects only the Sternberg classification Type is found and
is indicated in parenthesis; for SNe 1940E and 1963L the two sources disagree:
both are given;
$^b$ the offsets are in respect to the location of the supernova and are
in units of arcsec;
$^c$ ``indef'' indicates that the IRAF program ``phot'' did not calculate
the magnitude in some band and thus the color could not be derived;
the (--) indicates no images were taken in that band.
$^d$ limiting magnitudes corresponding to a $5 {\sigma}$ detection limit;
an asterisk next to the magnitude indicates that the magnitude corresponds
to a $3 {\sigma}$ detection limit: see text for details;
$^e$ standard deviations relative to the observed B, V and R magnitudes.

\medskip

TABLE 2b:

$^a$ The SN Type was taken from Branch 1992 and/or from the Sternberg Catalog;
``:'' after a Type indicates that according to
the Sternberg Catalog there is uncertainty; both Types are classified
spectroscopically, but the Type II are also classified according to the shape
of the light curve (P=plateau; L=linear) as given by the Sternberg Catalog.
This means that whenever a Type II is indicated as a Type IIP or IIL
the Sternberg classification was used;
for some objects only the Sternberg classification Type is found and
is indicated in parenthesis;
for SNEe 1961I, 1964F and 1967H the two sources disagree: both are given;
$^b$ the offsets are in respect to the location of the supernova and are
in units of arcsec.
$^c$ ``indef'' indicates that the IRAF program ``phot'' did not calculate
the magnitude in some band and thus the color could not be derived;
the (--) indicates no images were taken in that band; ``sat.'' for SN 1968L
in NGC 5236 indicates that the R band was saturated near the nucleus.
$^d$ limiting magnitudes corresponding to a $5 {\sigma}$ detection limit;
an asterisk next to the magnitude indicates that the magnitude corresponds
to a $3 {\sigma}$ detection limit: see text for details;
$^e$ standard deviations relative to the observed B, V and R magnitudes.

\medskip

TABLE 3

All data in this Table were derived from Tables 1a through 3b.

\medskip

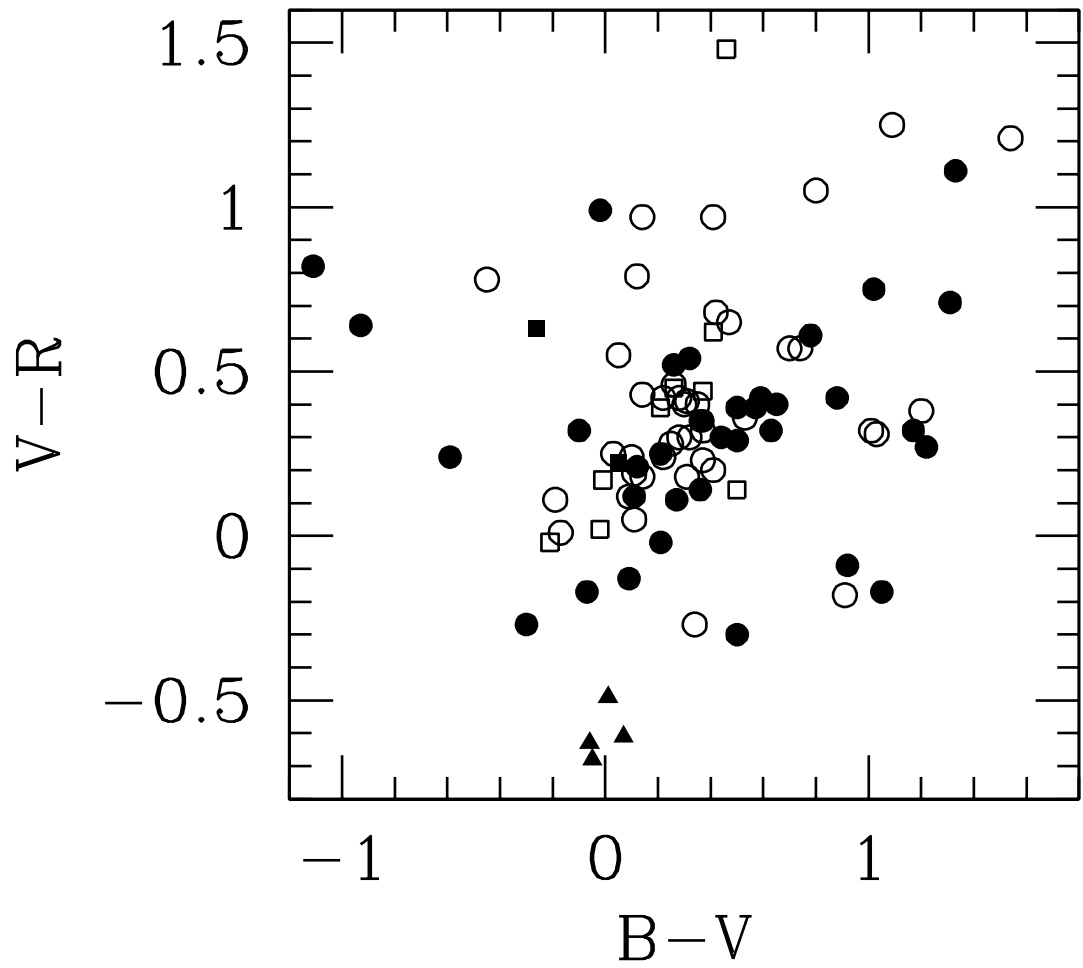
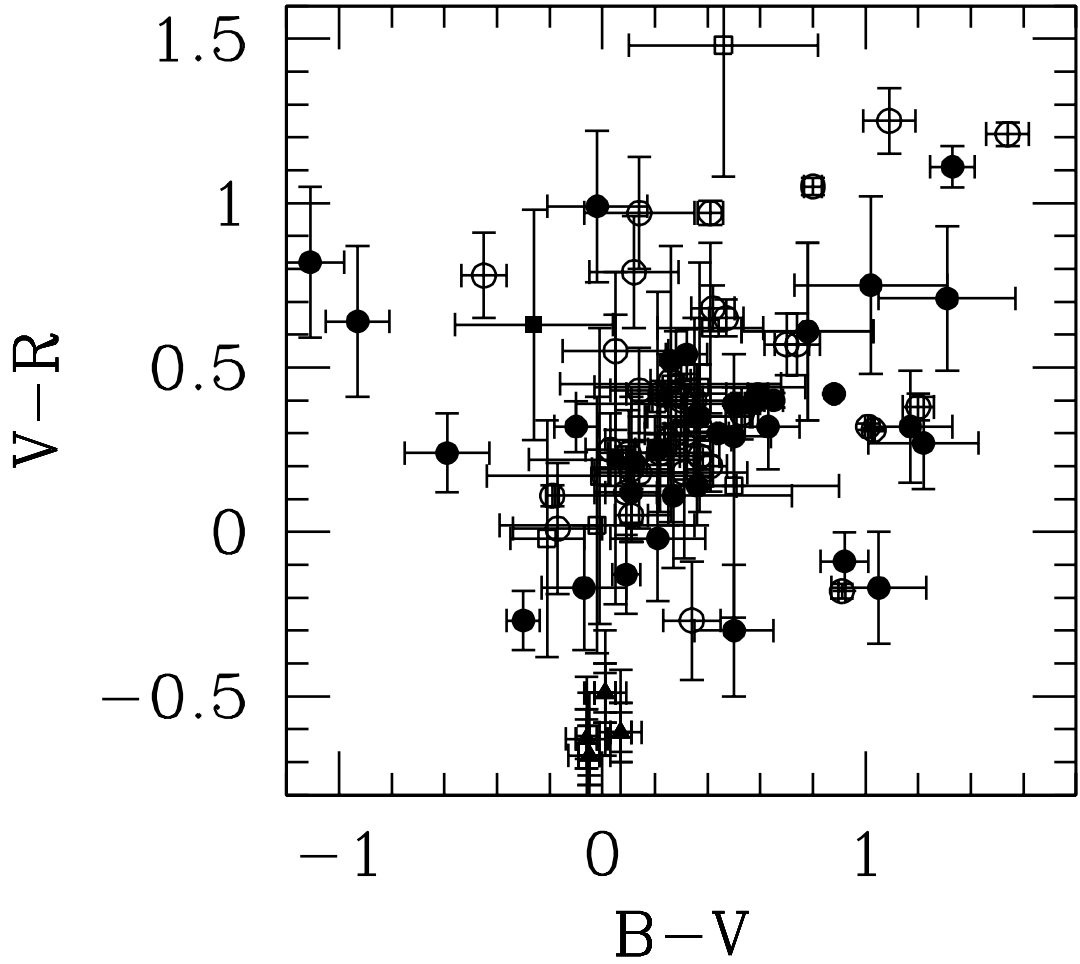
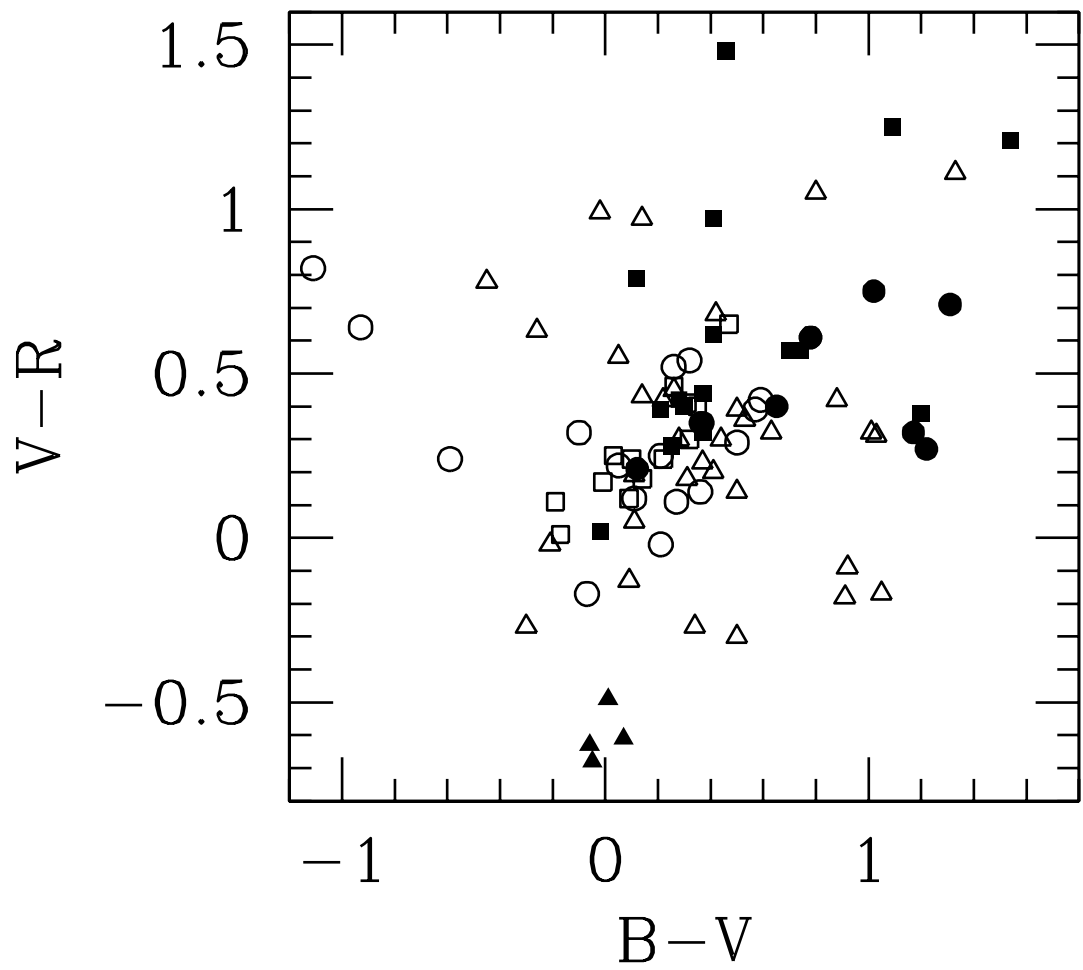
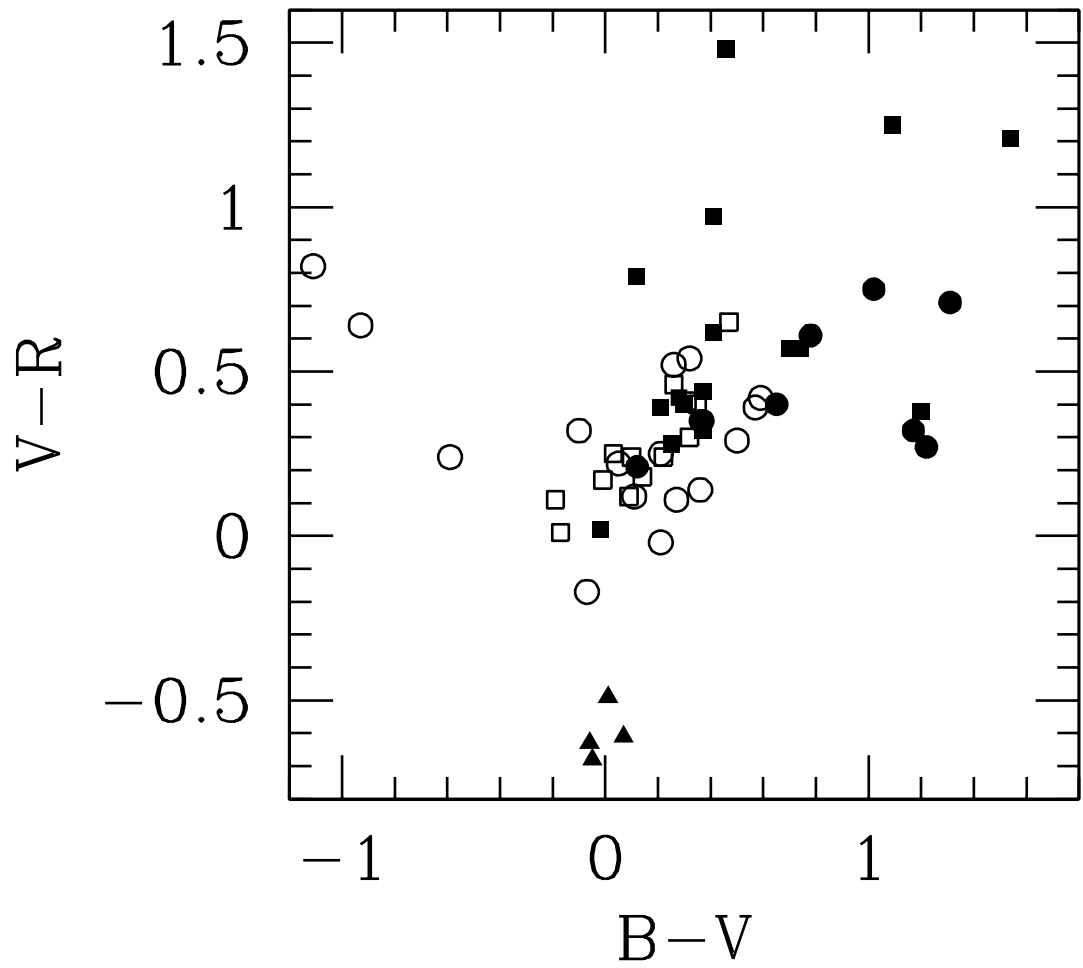
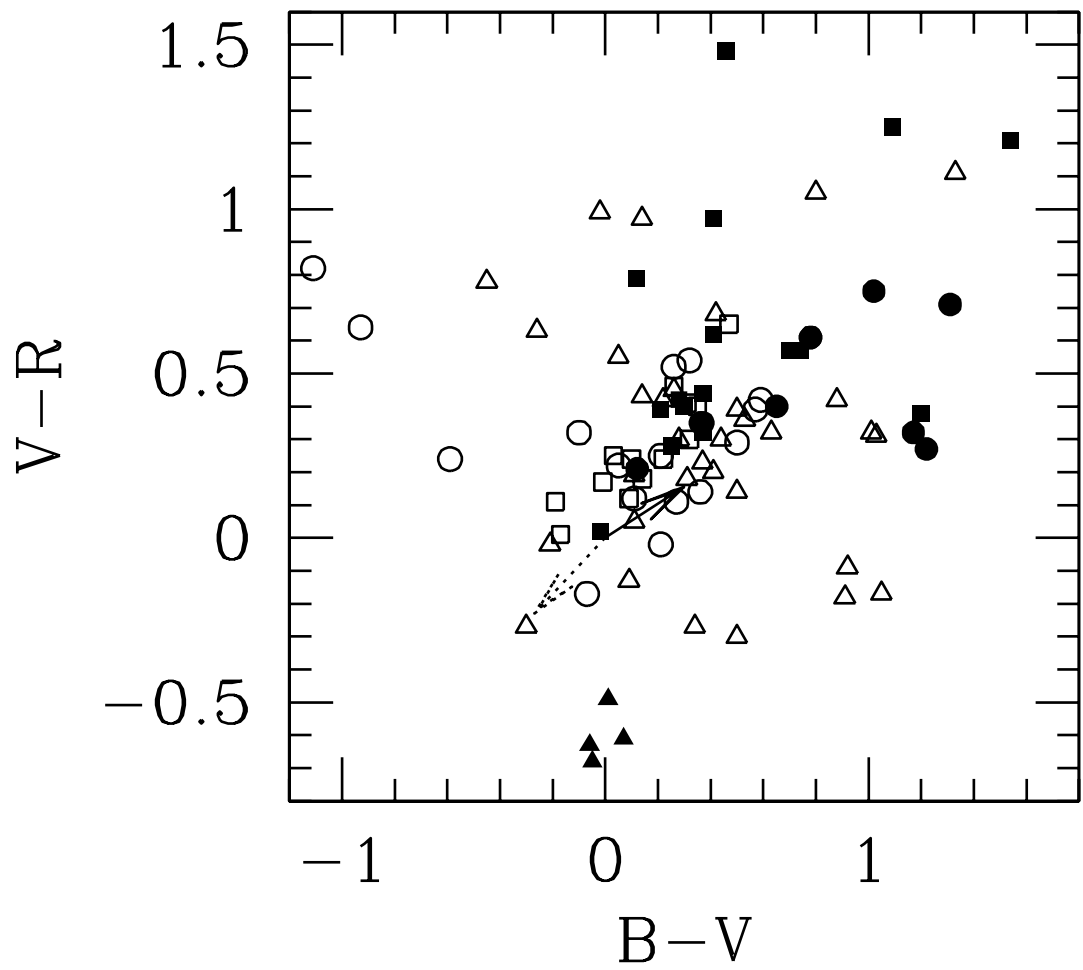
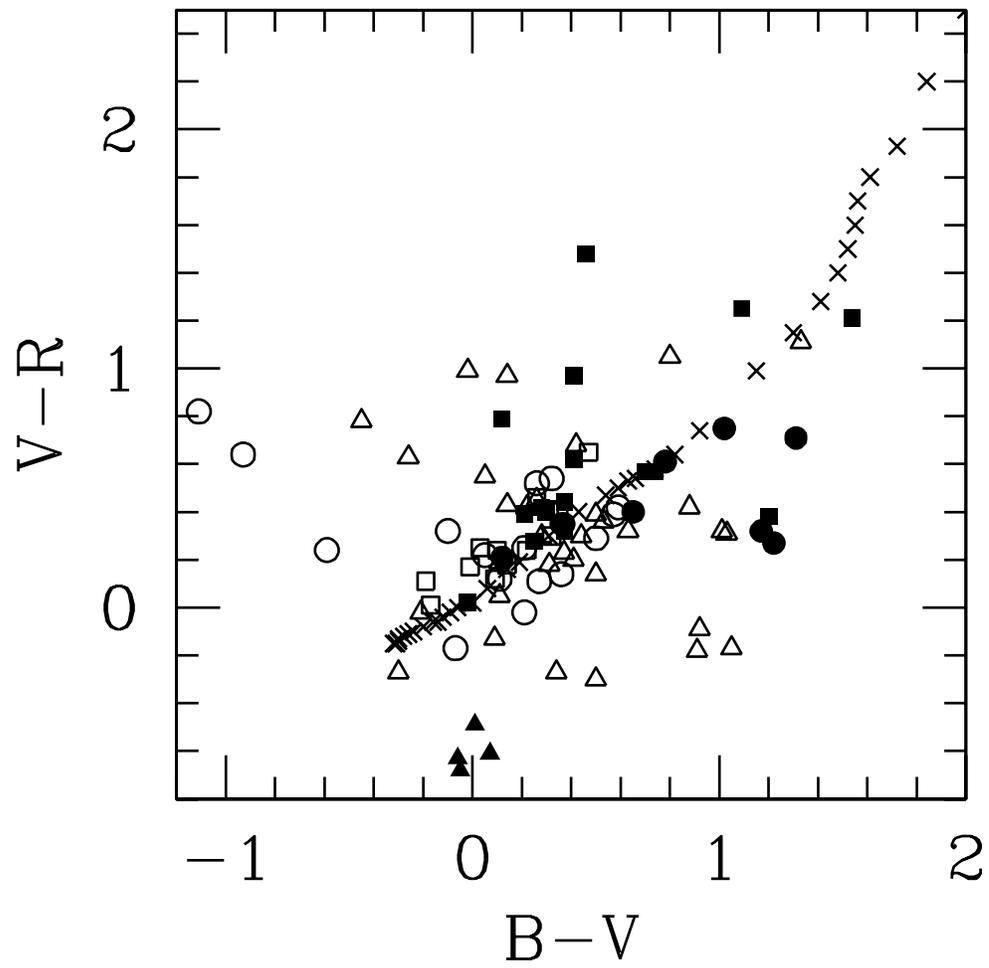
\end{document}